\documentclass[%
 aip,
 jmp,%
 amsmath,amssymb,
 reprint,%
 nofootinbib,%
 floatfix, %
]{revtex4-1}
\usepackage{graphicx}
\usepackage{caption}
\usepackage{subcaption}
\usepackage{dcolumn}
\usepackage{bm}
\usepackage[dvipsnames]{xcolor}
\usepackage{amssymb}
\usepackage{verbatim}
\usepackage{appendix}
\usepackage{listings}
\usepackage{units}
\usepackage{fancyhdr}
\usepackage{csquotes}
\usepackage{hyperref}

\usepackage{etoolbox}
\makeatletter
\patchcmd{\hyper@makecurrent}{%
    \ifx\Hy@param\Hy@chapterstring
        \let\Hy@param\Hy@chapapp
    \fi
}{%
    \iftoggle{inappendix}{
        \@checkappendixparam{chapter}%
        \@checkappendixparam{section}%
        \@checkappendixparam{subsection}%
        \@checkappendixparam{subsubsection}%
        \@checkappendixparam{paragraph}%
        \@checkappendixparam{subparagraph}%
    }{}%
}{}{\errmessage{failed to patch}}

\newcommand*{\@checkappendixparam}[1]{%
    \def\@checkappendixparamtmp{#1}%
    \ifx\Hy@param\@checkappendixparamtmp
        \let\Hy@param\Hy@appendixstring
    \fi
}
\makeatletter

\newtoggle{inappendix}
\togglefalse{inappendix}

\apptocmd{\appendix}{\toggletrue{inappendix}}{}{\errmessage{failed to patch}}
\apptocmd{\subappendices}{\toggletrue{inappendix}}{}{\errmessage{failed to patch}}

\newcommand*\obar[2][0.75]{
    \sbox{\myboxA}{$\m@th#2$}%
    \setbox\myboxB\null
    \ht\myboxB=\ht\myboxA%
    \dp\myboxB=\dp\myboxA%
    \wd\myboxB=#1\wd\myboxA
    \sbox\myboxB{$\m@th\overline{\copy\myboxB}$}
    \setlength\mylenA{\the\wd\myboxA}
    \addtolength\mylenA{-\the\wd\myboxB}%
    \ifdim\wd\myboxB<\wd\myboxA%
       \rlap{\hskip 0.5\mylenA\usebox\myboxB}{\usebox\myboxA}%
    \else
        \hskip -0.5\mylenA\rlap{\usebox\myboxA}{\hskip 0.5\mylenA\usebox\myboxB}%
    \fi}
\makeatother

\definecolor{Gold}{rgb}{1,0.84,0}

\definecolor{C1}{RGB}{51,34,136}
\definecolor{C2}{RGB}{136,204,238}
\definecolor{C3}{RGB}{68,170,153}
\definecolor{C4}{RGB}{17,119,51}
\definecolor{C5}{RGB}{153,153,51}
\definecolor{C6}{RGB}{221,204,119}
\definecolor{C7}{RGB}{102,17,0}
\definecolor{C8}{RGB}{204,102,119}
\definecolor{C9}{RGB}{136,34,85}
\definecolor{C10}{RGB}{170,68,153}

\definecolor{OC1}{RGB}{166,206,227}
\definecolor{OC2}{RGB}{31,120,180}
\definecolor{OC3}{RGB}{178,233,138}
\definecolor{OC4}{RGB}{51,160,44}
\definecolor{OC5}{RGB}{251,154,153}
\definecolor{OC6}{RGB}{227,26,28}
\definecolor{OC7}{RGB}{253,199,111}
\definecolor{OC8}{RGB}{255,127,0}
\definecolor{OC9}{RGB}{202,178,214}
\definecolor{OC10}{RGB}{106,61,154}

\definecolor{OOC1}{RGB}{195,170,60}
\definecolor{OOC2}{RGB}{93,54,134}
\definecolor{OOC3}{RGB}{101,188,103}
\definecolor{OOC4}{RGB}{194,106,187}
\definecolor{OOC5}{RGB}{112,143,57}
\definecolor{OOC6}{RGB}{108,126,215}
\definecolor{OOC7}{RGB}{182,120,55}
\definecolor{OOC8}{RGB}{70,193,154}
\definecolor{OOC9}{RGB}{185,74,115}
\definecolor{OOC10}{RGB}{186,76,65}

\definecolor{OOOC1}{RGB}{94,66,0}
\definecolor{OOOC2}{RGB}{71,103,222}
\definecolor{OOOC3}{RGB}{214,214,70}
\definecolor{OOOC4}{RGB}{85,0,71}
\definecolor{OOOC5}{RGB}{40,213,123}
\definecolor{OOOC6}{RGB}{220,46,88}
\definecolor{OOOC7}{RGB}{5,121,84}
\definecolor{OOOC8}{RGB}{255,169,246}
\definecolor{OOOC9}{RGB}{0,89,41}
\definecolor{OOOC10}{RGB}{250,143,56}

\DeclareMathOperator\erf{erf}

\newcommand{\p}{\ensuremath{\partial}}

\newcommand{\remove}[1]{{}}

\newcommand{\appref}[1]{\hyperref[#1]{Appendix~\ref{#1}}}

\usepackage{morefloats}

\newcommand{\phim}{\phi_{\rm max}}

\newcommand{\vb}{\bar{v}}
\newcommand{\xb}{\bar{x}}
\newcommand{\vh}{\hat{v}}

\newcommand{\vhpm}{\hat{v}_{\pm}}
\newcommand{\vhm}{\hat{v}_{-}}
\newcommand{\Vh}{\hat{V}}
\newcommand{\xh}{\hat{x}}

\newcommand{\Zj}{Z_j}

\newcommand{\mih}{\hat{m}_i}
\newcommand{\mjh}{\hat{m}_j}
\newcommand{\mzh}{\hat{m}_z}

\newcommand{\njh}{\hat{n}_j}

\newcommand{\fjh}{\hat{f}_j}

\newcommand{\Tih}{\hat{T}_i}
\newcommand{\Tjh}{\hat{T}_j}
\newcommand{\Tzh}{\hat{T}_z}

\newcommand{\phih}{\hat{\phi}}
\newcommand{\phimh}{\hat{\phi}_{\rm max}}
\newcommand{\gkeyll}{\texttt{Gkeyll}}

\begin{document}

\preprint{AIP/123-QED}

\title{Low Mach-number collisionless electrostatic shocks and
  associated ion acceleration} \author{I.~Pusztai}
\email{pusztai@chalmers.se} 
\affiliation{Department of Physics, Chalmers University of Technology,
  SE-41296 G\"{o}teborg, Sweden}
\author{J.~M.~TenBarge} 
\affiliation{Institute for Research in Electronics and Applied
  Physics, University of Maryland, College Park, MD 20742, USA}
\affiliation{Department of Astrophysical Sciences, Princeton
  University, Princeton, NJ 08543, USA} 
\affiliation{Princeton Plasma Physics Laboratory, Princeton, NJ 08543, USA}
\author{A.~N.~Csap\'o} 
\affiliation{Department of Physics, Chalmers University of Technology,
  SE-41296 G\"{o}teborg, Sweden}
\author{J.~Juno} 
\affiliation{Institute for Research in Electronics and Applied
  Physics, University of Maryland, College Park, MD 20742, USA}
\author{A.~Hakim} 
\affiliation{Princeton Plasma Physics Laboratory, Princeton, NJ 08543, USA}
\author{L.~Yi} 
\affiliation{Department of Physics, Chalmers University of Technology,
  SE-41296 G\"{o}teborg, Sweden}
\author{T.~F\"ul\"op} 
\affiliation{Department of Physics, Chalmers University of Technology,
  SE-41296 G\"{o}teborg, Sweden}

\date{\today}

\begin{abstract}
  The existence and properties of low Mach-number ($M \gtrsim
    1$) electrostatic collisionless shocks are investigated with a
  semi-analytical solution for the shock structure. We show that the
  properties of the shock obtained in the semi-analytical model can be
  well reproduced in fully kinetic Eulerian Vlasov-Poisson
  simulations, where the shock is generated by the decay of an initial
  density discontinuity. Using this semi-analytical model, we study
  the effect of electron-to-ion temperature ratio and presence of
  impurities on both the maximum shock potential and Mach number.  We
  find that even a small amount of impurities can influence the shock
  properties significantly, including the reflected light ion
  fraction, which can change several orders of magnitude.
  Electrostatic shocks in heavy ion plasmas reflect most of the
  hydrogen impurity ions.
\end{abstract}

\keywords{collisionless shock, ion acceleration, laser plasma}

\maketitle

\section{Introduction}
\label{sec:intro}
Collisionless shocks are common in space, astrophysical and laboratory
plasmas, and their efficiency as particle accelerators is well
established \cite{tidmankrall,malkowith}. 
In the context of laser-produced plasmas, collisionless shocks
may be used for ion acceleration.  When a laser hits an over-dense
plasma, it leads to electron heating and density steepening. This scenario can
result in a collisionless shock, which propagates into the target.
Ions can be reflected off the moving electrostatic potential front
with twice the shock velocity in the rest frame of the upstream
population. Recent experimental and numerical results have shown that
mono-energetic acceleration of protons can be achieved at modest laser
intensities, albeit with rather low laser-to-particle energy
conversion efficiency \cite{Haberberger2012,fiuza,BSW}.  High energy
ions with narrow energy-spread would be very attractive for a wide
range of applications.

In particular, non-relativistic electrostatic collisionless shocks
have been observed both in the laboratory
\cite{taylor,romagnani,morita} and in space \cite{mozer} and can also
be of interest in laser-plasma driven shock acceleration of
protons. The theoretical basis for electrostatic,
  ion-acoustic-type shocks were laid out already in the 1960s
  \cite{sagdeevRev66}, and it was realized that such shocks can be
  sustained by the reflection of ions on the shock potential
  \cite{Moiseev63}. Numerical studies of collisionless shocks
  \cite{Forslund,dieckmann2010,sarri2010,macchi12,fiuza} have mostly used
  Particle-In-Cell \cite{simulationmethods} algorithms. Eulerian
  Vlasov-Maxwell approaches avoid issues with particle statistics;
  this difference can be particularly important for the accurate
  modeling of scenarios where the reflected ion fraction is small, or
  where turbulent fluctuations shape the shock dynamics
  \cite{grassi16}. However, there are only a limited number of
  electrostatic shock studies using Vlasov-Maxwell tools
  \cite{BSW,grassi16}.

Analytical models are appealing due to their simplicity and can
  be useful for gaining physical understanding as a complement to more
  extensive numerical simulations. Analytical models of various
  sophistication range from cold ion fluid approaches \cite{Moiseev63}
  to exact kinetic shock solutions \cite{smith70}. A simple treatment
  of the non-magnetized collisionless shock structure, taking into
  account finite ion temperature, is given in
  Refs.~\citenum{cairns} and \citenum{CairnsPPCF}. In the current
  paper, we adopt this formalism but with a more consistent treatment
  of the trapped regions in the ion phase space, to address the
  effects of the ion composition on shock properties and the reflected
  ion fraction in low Mach number (i.e., $M\approx 1-1.5$) electrostatic
  shocks.

  First, through comparisons to fully kinetic simulations using
     the Eulerian Vlasov-Poisson solver contained in
       \gkeyll~\cite{juno}, we demonstrate that the properties of the
     shock are well reproduced by the semi-analytical model.  Then,
   we use the model to investigate the effect of a heavy ion
   component on the existence of the shock, the Mach number, and
   reflected ion fraction.  We show that the effect of a heavy ion
   component can be important also if it is only present in small
   quantities, as it may affect the electrostatic potential and the
   shock propagation speed, and thereby have a strong influence on the
   ion spectrum. For instance, only a few percent of a carbon impurity
   increases the maximum electrostatic potential significantly for the
   same electron-to-ion temperature ratio and Mach number. This
   potential, in turn, affects the reflected ion fraction
   exponentially. Even the existence of shock solutions is
   affected by the impurity concentration.

We have also investigated shock properties in the case where the main ion
component is a heavy ion and the trace impurity is hydrogen. This case is
the typical scenario in laser-driven ion acceleration
experiments, where solid aluminum foil target having hydrogen
  impurities on their surface is often used. In this case, the
semi-analytical model predicts that the hydrogen ions are almost all
reflected. This prediction is corroborated with Vlasov-Poisson
simulations, showing excellent agreement both with the distribution
functions and the shock speed.

The structure of the paper is the following. In Section
\ref{sec:secone}, we describe the semi-analytical model and the
assumptions behind it. Next, in Section \ref{sec:sectwo} we
compare the model to kinetic simulation results using a
Vlasov-Poisson solver, and demonstrate the similarities of the ion and
electron distribution function, the electrostatic potential and Mach
number. In Section \ref{sec:secthree}, we proceed to use the
semi-analytical model to study the effect of ion composition and
electron-to-ion-temperature ratio. Finally, we summarize and conclude
in Section~\ref{sec:secfour}.

\section{Electrostatic shock}
\label{sec:secone}
Electrostatic shocks are sustained by an electric field that is linked
to a density gradient between the downstream and upstream
plasmas. We focus on the effect of ion composition on the shock
  properties, while we are not concerned about which physical process
  generated the shock, or about its long term stability. In the
  following, we describe a simple, one-dimensional model for a
  collisionless electrostatic shock, which is sufficient for our
  purposes, and it is similar to the model of Ref.~\citenum{cairns}.

\begin{figure*}
\includegraphics[width=0.35\textwidth]{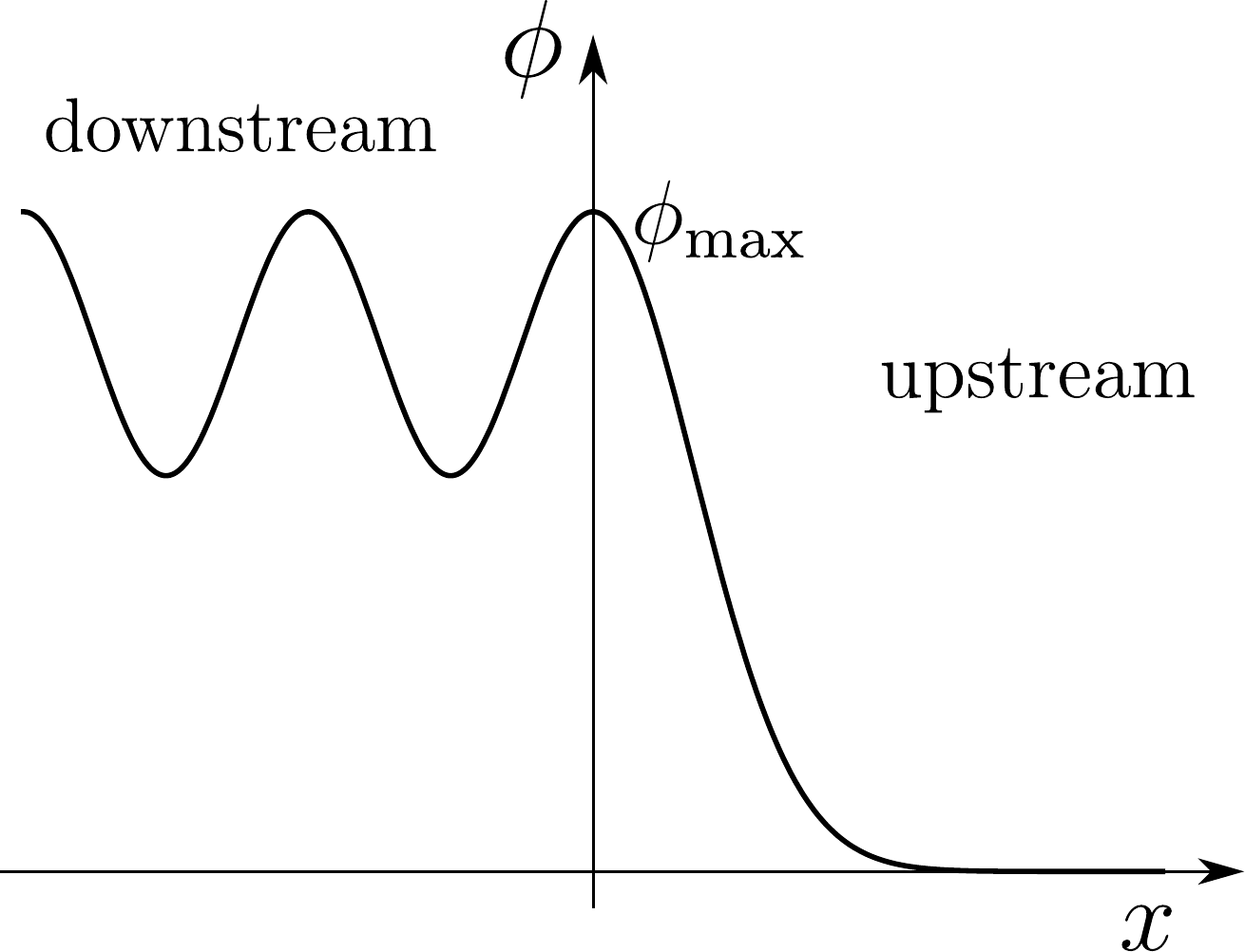}
\put(-140,115){\Large (a)}
\hspace{10mm}
\includegraphics[width=0.35\textwidth]{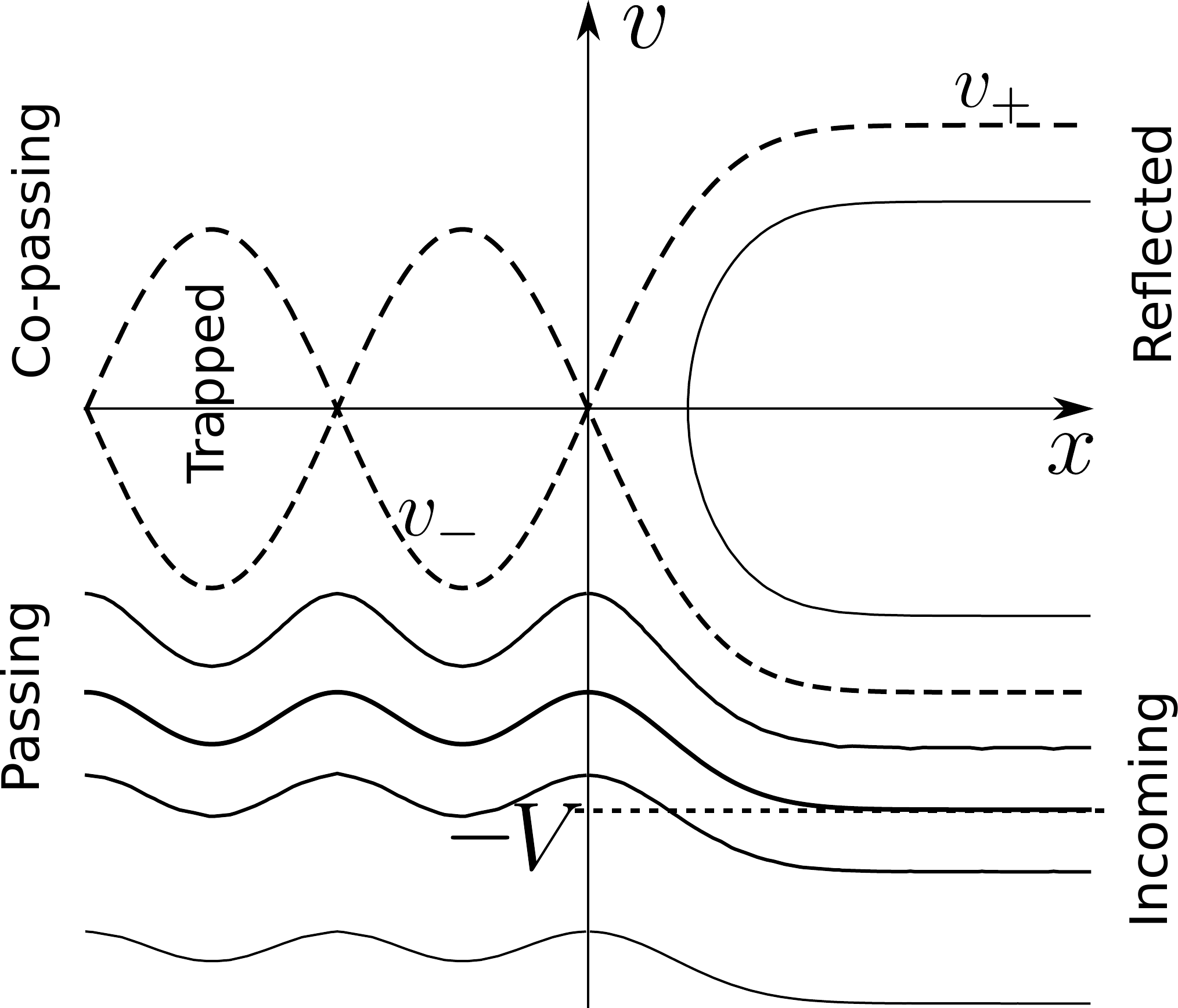}
\put(-140,115){\Large (b)}
\caption{\label{sketch} (a) Electrostatic potential of the shock
  structure showing a monotonic increase from $\phi=0$ to $\phim$ in
  the upstream region, and an oscillatory behavior with
  $0<\phi\le\phim$ in the downstream region. (b) Phase-space plot of
  the ion distribution showing the different populations (incoming,
  reflected, passing). In the upstream region the shock potential
  reflects a fraction of the ions, while in the downstream the density
  of passing ions is oscillatory in $x$.}
\end{figure*}

The potential $\phi$ increases from zero in the far upstream region
($x\rightarrow \infty $) to some positive value $\phim$ at $x=0$, as
illustrated in Fig.~\ref{sketch}a. In the downstream region,
  $x<0$, the electrostatic potential is finite and oscillatory. In the
laboratory frame, the shock propagates in the $+x$ direction with a
velocity $V$, whereas the ions sufficiently far ahead of the shock
have a zero flow speed. We assume $V\sim c_s\gg v_i$, where
$v_i=\sqrt{T_i/m_i}$ is the thermal speed of the main ions with mass
$m_i$ and temperature $T_i$, and $c_s=\sqrt{Z_iT_e/m_i}$ is the sound
speed, with $T_e$ the electron temperature and $Z_i$ the ion charge
number.  Thus, in the shock frame, the far upstream ion flow
  velocity is $-V$; these ions represent the incoming population in
Fig.~\ref{sketch}b. The phase-space separatrix between ions that
  can pass through the potential barrier and those reflected from it
  is represented by the dashed line in the $x>0$ and $v<0$ region of
  Fig.~\ref{sketch}b. We consider only a certain vicinity near the
shock, where the reflected ions are already present for $x>0$, and do
not treat the problem of what happens with the leading edge of the
reflected population (i.e.,~the so-called ``foot'' problem
  \cite{malkov}).

We use a notation and normalizations that accommodate multiple ion
species with arbitrary charge and mass. The ``bulk'' ion species
that dominates the dynamics will be denoted by index $i$, and the
impurities by index $z$, while $e$ refers to electrons and $j=\{i,\,
z\}$ to ions in general. For any physical quantity $X$, we define a
dimensionless normalized quantity $\hat{X}=X/\bar{X}$, with $\bar{X}$
a species-independent normalizing quantity. In particular
$\bar{T}=T_i$ is the bulk ion temperature, $\bar{n}=n_0$ is the far
upstream incoming bulk ion density, $\bar{\phi}=T_i/e$,
$\bar{v}=\sqrt{T_i/m_p}$ with the proton mass $m_p$,
$\xb=[T_i\epsilon_0/(e^2n_0)]^{1/2}$ (which is $\xb=\vb/\omega_{pp}$,
with $\omega_{pp}$ the proton plasma frequency at
$n_0$). Finally, distribution functions are normalized by
  $\bar{f}=\bar{n}/\bar{v}$.

The general
solution of the collisionless, steady state ion kinetic equation
\begin{equation}
v \frac{\p f_j}{\p x}-\frac{Z_je}{m_j}\frac{d\phi}{d x}\frac{\p f_j}{\p v}=0
\end{equation} 
is $f_j=f_j(E_j)$, with $E_j=m_jv^2/2+Z_je\phi$ the total energy.
Thus, the distribution prescribed at the boundaries where particles
are incoming into our domain ($v<0$ at $x\rightarrow \infty$ and $v>0$
at $x \rightarrow -\infty $) simply needs to be extended along the
contours of constant total energy. Such lines are shown by solid lines
in Fig.~\ref{sketch}b, which then represent contours of the
distribution function, with the thick line corresponding to the
maximum of the distribution. We take the far upstream ion distribution
to be a Maxwellian,
$f_j^{(+\infty)}=n_j/(v_j\sqrt{2\pi})\exp\left[-(v+V)^2/(2v_j^2)
  \right]$, for $x\rightarrow \infty$ and $v<0$. We are also free to
specify the ion distribution in the \emph{trapped} regions of
phase-space (see Fig.~\ref{sketch}b); we assume that the trapped
regions are empty. Furthermore, consistent with the assumption
$V>v_i$, we neglect the small \emph{co-passing} population, appearing
above the trapped and the reflected regions, as it would represent a
tail of a Maxwellian and its contribution to a charge imbalance
between upstream and downstream is negligible.

Thus, the normalized ion distribution function and the ion density are given by
\begin{align}
\njh^\pm(\xh) & = \int_{-\infty}^{\vhpm} \fjh d\vh
\equiv\frac{\njh}{\sqrt{2\pi\Tjh/\mjh}} \nonumber \\ \times
\int_{-\infty}^{\vhpm} & d\vh\; \exp
\left\{-\frac{\left(\sqrt{\vh^2+2\Zj\phih/\mjh}-\Vh\right)^2}{2
  \Tjh/\mjh} \right\},\label{nitotal}
\end{align}
where the $+$ and $-$ upper indices refer to upstream and
  downstream, respectively. Above the velocity separatrices given by
  $\vhpm=\pm \sqrt{2(\Zj/\mjh)(\phimh-\phih)}$, the distribution
  function vanishes: $\fjh(\vh>\vhpm)=0$ for $\pm \xh>0$. In the far
upstream region (i.e., where the reflected population is present,
  but $\phih=0$) the density can be explicitly evaluated to
\begin{equation}
  \njh^+(+\infty) =  \frac{\njh}{2}\left[ 1+2
    \erf\left(\tilde{V}_j\right) +  \erf
    \left(\sqrt{\Psi_j}-\tilde{V}_j\right) \right],
\label{ni}
\end{equation}
where $\tilde{V}_j=\Vh/\sqrt{2\Tjh/\mjh} $,
  $\Psi_j=Z_j\phimh/\Tjh$, and $\rm \erf$ denotes the Gauss error
  function.

The ion density in Eq.~(\ref{nitotal}) is similar to the one used in
Ref.~\citenum{cairns}, except that there the distribution
function appears to be extended up to $\vh=0$ in the downstream
region, even when $\phih<\phimh$. This assumption is not consistent
with the constancy of the distribution along the energy contours in a
steady state. Although the incorrect treatment does not effect
$\phimh$, because that depends on the upstream distribution function,
it increases the amplitude and the wavelength of the oscillation of
$\phih$ in the downstream region.

The reflected ion fraction can also be calculated from the ion
distribution function:
\begin{align}
  \alpha_j= &\frac{\int_{0}^{\sqrt{2\phimh Z_j/\mjh}} d\vh\; \exp
    \left\{-\frac{(\vh-\Vh)^2}{2 \Tjh/\mjh} \right\}}
        {\int_{-\infty}^{0} d\vh\; \exp \left\{-\frac{(\vh+\Vh)^2}{2
            \Tjh/\mjh} \right\}} \label{alpha} \\ = & \left[
          \erf\left(\tilde{V}_j \right)+ \erf
          \left(\sqrt{\Psi_j}-\tilde{V}_j\right)\right]\left/\left[
          1+\erf\left(\tilde{V}_j\right) \right].\right. \nonumber
\end{align}

Electron distribution functions assuming adiabatic trapping throughout
the $m_e v^2/2<e \phi_{\rm max}$ region of phase space have been
considered previously \cite{schamel72,dudnikova16}. Even though
trapping can occur in the cases that we will consider, it is
restricted to the regions bounded by the local potential minima of the
downstream oscillation: $m_e v^2/(2e) < \phi_{\rm max}-\phi_{\rm
  min,-1}$, where $\phi_{\rm min, -1}$ denotes a local potential
minimum. Note, that the electrons that are \emph{not} trapped in these
downstream oscillations, but are merely constrained to the
semi-infinite downstream region (i.e., those with $\phi_{\rm
  max}-\phi_{\rm min,-1}<m_e v^2/(2e)< \phi_{\rm max} $), have an
infinitely long bounce time (unlike trapped particles considered in
Ref.~\citenum{gurevich}), and as such, they do not behave as real
trapped particles. For simplicity, we will neglect a possible
flattening of the distribution in the real trapped regions and
consider a Maxwell-Boltzmann electron distribution $f_e=
n_e/(v_e\sqrt{2\pi})\exp\left[-v^2/(2v_e^2)+ e\phi/T_e \right]$, where
$v_e=\sqrt{T_e/m_e}$ is the electron thermal speed, and the constant
$n_e$ is the far upstream electron density.  As the flow speed $V/v_e$
is small in the limit $\sqrt{m_e/m_i}\ll 1$, the flow of the electron
distribution in the shock frame is neglected. Thus, the normalized
electron density is $\hat{n}_e(\xh)=\hat{n}_e e^{\phih/\tau}$, where
$\tau= T_e/T_i(=\hat{T}_e)$.  To obtain the constant $\hat{n}_e$ we
may assume that sufficiently far upstream the plasma is
quasineutral. Thus the electron density is
\begin{equation}
\hat{n}_e(x)=\sum_j Z_j\njh^{+}(+\infty)e^{\phih/\tau},
\label{qnadi}
\end{equation} 
 where $\njh^{+}(+\infty)$ is given by Eq.~(\ref{ni}).

Finally, the ion and electron densities can be used together with
Poisson's equation to find the electrostatic field. In normalized
quantities, Poisson's equation reads
\begin{equation}
\frac{d^2\phih(\xh)}{d\xh^2}=\hat{n}_e(\xh) - \sum_jZ_j\hat{n}_j(\xh).
\label{poisson2}
\end{equation} 

Following the classic treatment \cite{tidmankrall}, Poisson's equation
can be rewritten in terms of a Sagdeev potential $\Phi(\phih,
\phimh)=\int_0^{\phih}\left[
  \sum_j \Zj \njh (\phih',\phimh)-\hat{n}_e(\phih', \phimh) \right] d
\phih' $ so that we have $ d^2\phih/d \xh^2=-\partial \Phi/\partial
\phih $ that, after multiplication by $d \phih/d\xh$, gives the
familiar equation
\begin{equation}
\frac{1}{2}\left(\frac{d\phih}{d\xh}\right)^2=-\Phi,
\label{sagdeev}
\end{equation}
analogous to the equation of motion of a particle in a potential. The
condition $\Phi(\phimh, \phimh)= 0$ determines the quantity
$\phimh$. A solitary wave occurs when the Sagdeev potential $\Phi$ has
a local maximum at the origin ($\phih=0$) and goes through zero again
at some finite value of $\phih $. Shock-like structures can form when
there is damping in the system, for example ion-reflection, which
produces an asymmetry between upstream and downstream sides.

For a laser with moderately high intensity, the electron
  temperature can be expected to be in the $\rm MeV$ range
  \cite{wilks,fiuza}, and $\tau$ can range from a few tens to several
  hundred. Although the dispersion relation of sound waves in
  multi-species plasmas can be rather complex \cite{burton71}, for
  simplicity, we define the Mach number $M=V/c_s=\Vh\sqrt{\mih/(\tau
    Z_i)}$, with respect to the speed $c_s=\sqrt{Z_iT_e/m_i}$. 

In the model used here -- similarly to Ref.~\citenum{cairns} -- both
$\tau$ and $\Vh$ (or equivalently the Mach number $M$) are treated as
inputs. For a given $\tau$ there can be a finite range of Mach numbers
where shock solutions exist. Towards the highest Mach number of such a
range, the amplitude of the downstream oscillation approaches zero (with
decreasing wave length). The degenerate case of the amplitude becoming
zero corresponds to a \emph{monotonic shock} structure.  The boundary
condition $\partial \Phi/\partial \phih=0$ at $\phih=\phimh$ produces
such monotonic shock solutions. Representing an additional constraint,
this boundary condition removes one degree of freedom, thus it can be
used to calculate $\Vh$ for a given $\tau$, as done in
Ref.~\citenum{eliasson}.

\section{Comparison to kinetic simulations}
\label{sec:sectwo}
We have performed simulations with the \gkeyll~\cite{juno}
Vlasov-Poisson solver and compared the results to the analytical model
presented in the previous section. The simulations evolve kinetic ion
and electron species, starting from a density step as an initial
condition. We focus on the electrostatic shock that develops at the
initial density discontinuity and propagates into the low density
region. For the comparisons, we use 1x1v (one spatial and one
  velocity dimension), noting that increasing the dimensionality to
  2x2v leads to identical results.\footnote{This behavior is
    expected, as in the absence of initial magnetic perturbations, a
    1x1v problem remains 1x1v in continuum Vlasov simulations, since
    no coupling occurs to other velocity dimensions. This behavior is
    unlike PIC codes, where the statistical noise can provide a seed
    to various instabilities \cite{kato,dieckmann13,stockem14} that
    can potentially break the initial symmetry of the solution. However, we
    note that for the setup studied here, no growth of instabilities
    was observed in a 2x3v PIC simulation either, using the EPOCH code
    \cite{epoch}; in particular, the noise in the magnetic field stayed
    at a constant level throughout the simulation.}

The possible range of Mach numbers where shock solutions exist in our
model increases with the temperature ratio $\tau$. In particular,
below a certain $\tau$ value, no solutions exist. Although our ion
distribution function is slightly different from that of
Ref.~\citenum{CairnsPPCF}, in a single species plasma we obtain a
similar result for the allowed Mach number range to their
Figure~2. The reason for this insensitivity to the exact form of the
distribution is the following. At the upper Mach number
  threshold, where the monotonic shocks appear, $\vhm=0$ thus the two
types of distribution function coincide. At the lower Mach number
  threshold, $\phimh$ (determined by upstream dynamics) goes to zero,
and the reflected fraction diminishes, as we will show. When this
happens, the population above $\vh_-$ that distinguishes the two
distribution functions, becomes negligibly small, and thus does not
affect quasineutrality.

Therefore, we find it useful to choose a moderate value of $\tau$ with
a wide range of possible solutions. This choice is convenient, since
unlike in the analytical model, where the inputs are $\Vh$ and $\tau$,
and $\phimh$ is an output, in the simulation the inputs are the ratio
of densities in the density discontinuity, and $\tau$. In this case,
the density ratio determines the total potential drop across the whole
simulation domain -- as the electron distribution is close to
Maxwell-Boltzmann -- and the potential drop across the shock
(i.e.,~$\phimh$) is approximately half of the total potential drop. By
adjusting the density ratio, the shock potential and the shock
propagation speed can be adjusted.

In the following comparison we assume $\tau=45$, an initial density
ratio $n_i(x=-\infty)/n_0=1.5$, and $v_e=0.1 c$. Both species are
kinetic, with a mass ratio $m_i/m_e=1836$, the initial discontinuity
in the density is located in the middle of the $L_x=100 \lambda_D$
wide simulation domain that has $768$ grid points
($\lambda_D=\sqrt{\epsilon_0T_e/(e^2n_e)}$ is the Debye-length). The
range of electron velocities is $[-6 v_e,6v_e]$ and for ions it is
$[-6 v_i,12v_i]$. The number of cells is 256 and 64 in configuration
space and velocity space, respectively, with polynomial order 2. The
boundary conditions are open in configuration space and zero flux in
velocity. 

\begin{figure}
\includegraphics[width=0.45\textwidth]{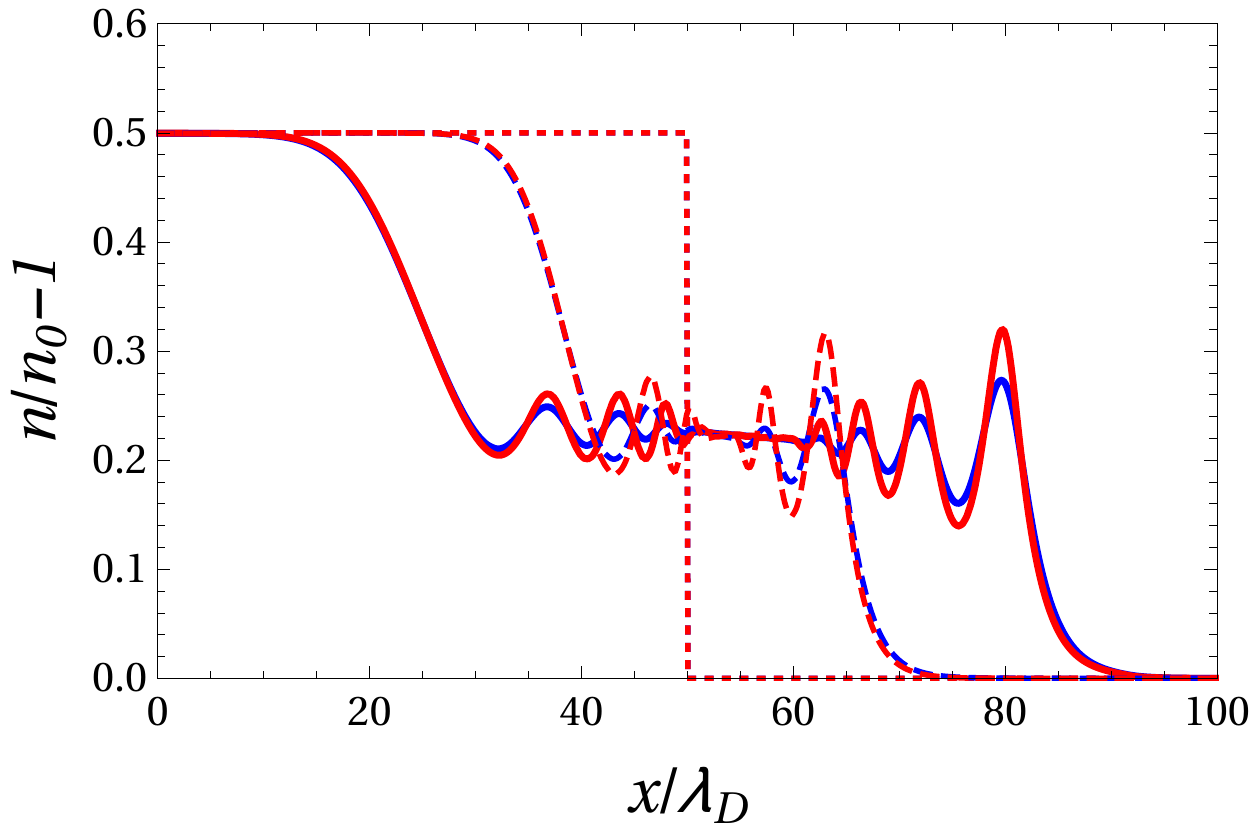}
\caption{\label{densityex} Relative density difference from the far
  upstream value. Red: ion; blue: electron. Dotted lines: $t
  \omega_{pp}=0$, dashed lines: $t \omega_{pp}=15$, thick solid lines:
  $t \omega_{pp}=30$.}
\end{figure}

The relative density difference from the far upstream value ($n_0$)
for both electrons (blue curves) and ions (red curves) is shown in
Fig.~\ref{densityex} at $t\omega_{pp}=\{0,\,15,\,30\}$. There are
features of the shock in the simulation, related to the initial
condition, which are not present in the analytical model: the
downstream oscillatory part of the solution decays away from the shock
front, similarly to the ion acoustic front structures found
  analytically in Ref.~\citenum{mason70} for small density
  discontinuity initial conditions. Further downstream, the solution
transitions into another oscillation, which leads up to the
rarefaction front propagating in the opposite direction.  Also,
considering the distance between the initial density step location
($x/\lambda_D=50$) and the maximum density point in the shock at
$t\omega_{pp}=15$ and $30$ we find that the propagation speed is
increasing with time. Nevertheless, some properties of the shock front
become approximately independent of the initial condition during the
simulation. In particular the maximum value of the shock densities and
their shape in the upstream region do not change significantly between
$t\omega_{pp}=15$ and $t\omega_{pp}=30$.

\begin{figure*}
\includegraphics[width=0.44\textwidth]{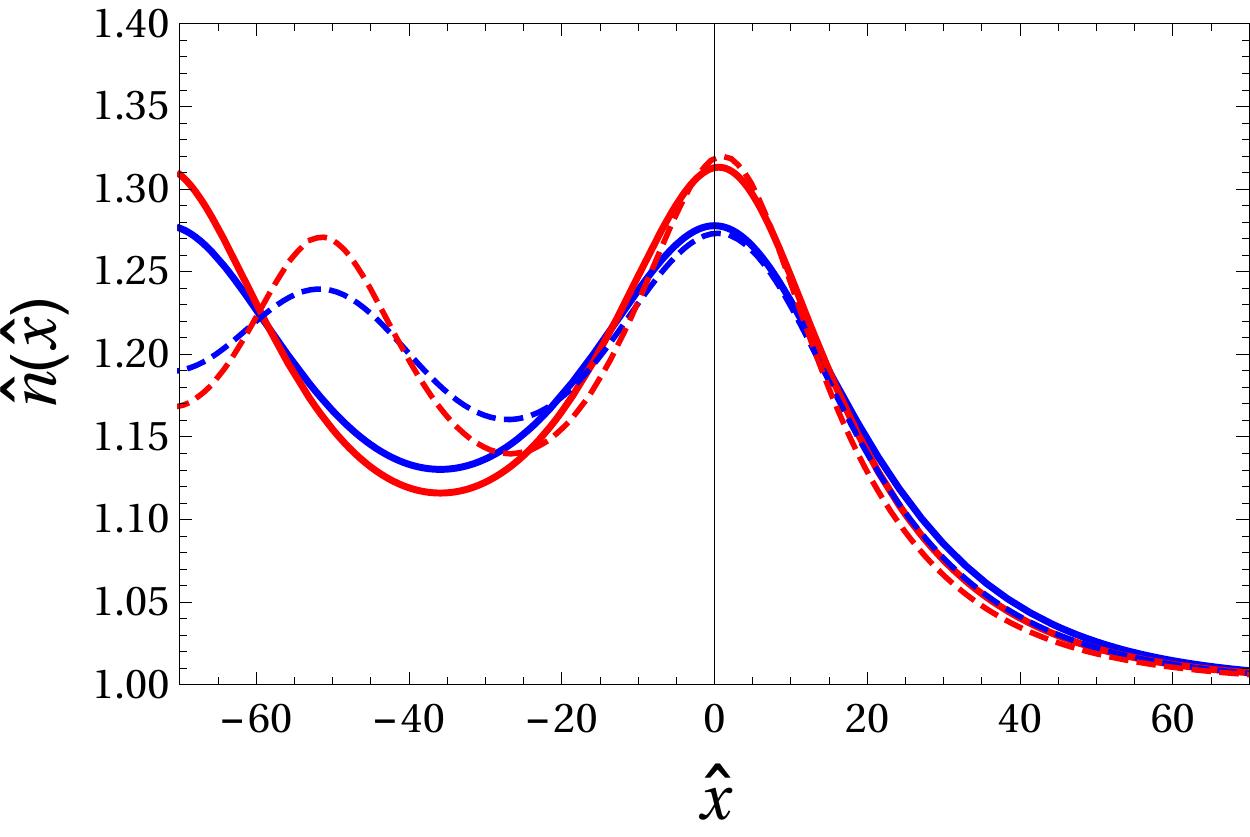}
\put(-30,95){\Large (a)}  
\put(160,95){\Large (b)}
$\quad$
\includegraphics[width=0.45\textwidth]{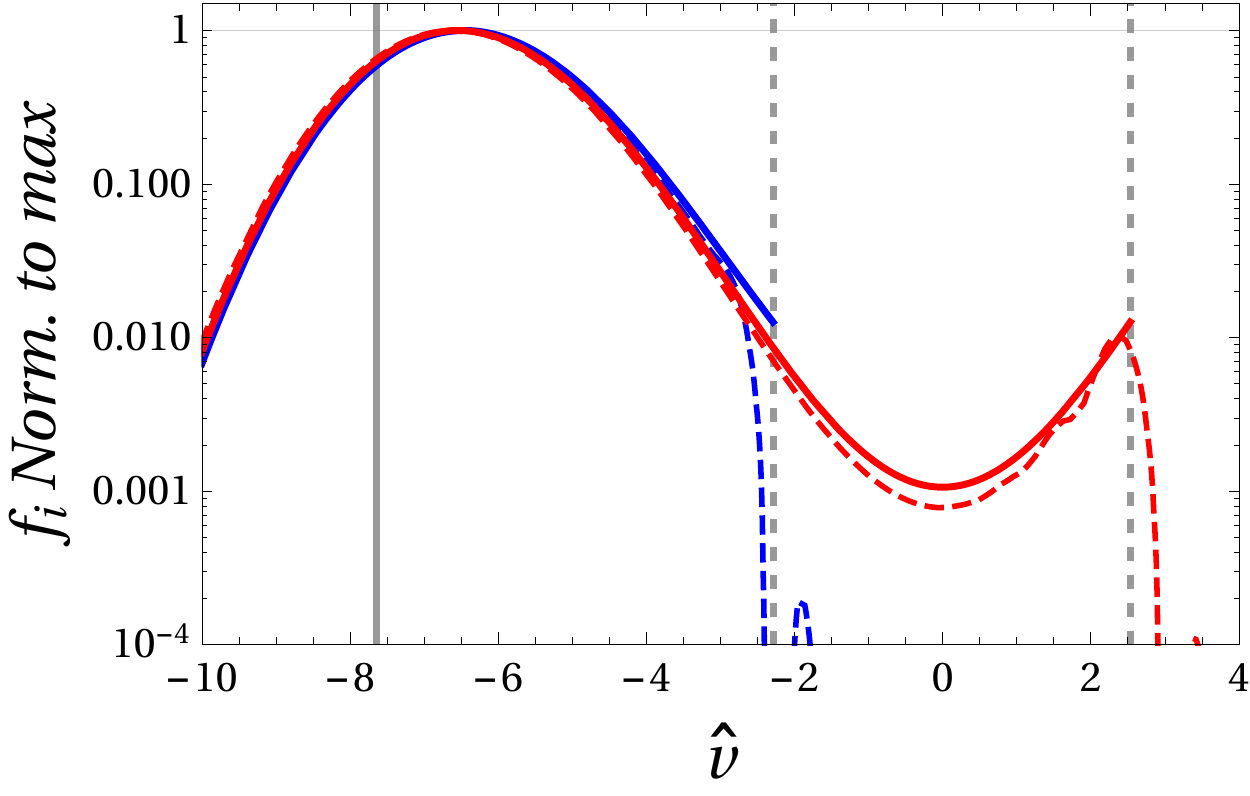}
\caption{\label{densities} Comparison of analytical and simulation
  results. (a) Relative density difference from the far upstream
  value. Red: ion; blue: electron. Thick solid lines: analytical
  model; dashed lines: simulation at $t \omega_{pp}=30$. $\xh$ denotes
  the normalized length from the top of the shock. (b) The ion
  distribution function $f_i$ normalized to its maximum value in the
  upstream ($\hat{x}=15$, red) and downstream ($\hat{x}=-15$, blue)
  regions. Thick solid lines: analytical; dashed lines: simulation
  result at $t \omega_{pp}=30$.  Solid vertical bar marks the far
  upstream ion flow speed in shock frame ($-\Vh$), and dotted vertical
  bars correspond to $\vh_-$ and $\vh_+$.  }
\end{figure*}

Let us compare the simulation result to the analytical model at
$t \omega_{pp}=30$. Using the analytical model for $\tau=45$, we
identified a $V$ that produces similar maximum ion and electron
densities to those in the simulated shock front. The result shown in
Fig.~\ref{densities}a with solid lines correspond to $\Vh=7.64$
($M=1.139$), with $\phimh=10.96$. The simulation result is shown with
dashed lines, and $\hat{n}_e$ ($\hat{n}_i$) corresponds to blue (red) curves.
The propagation speed is slightly higher than the one estimated for
the simulation $M=1.133\pm 0.003$ from the movement of the highest
density position, although as we mentioned, the propagation speed has an
increasing trend during the simulation. We can see that the difference
between the maximum $\hat{n}_i$ and $\hat{n}_e$ values is very well captured.

Next, we compare the upstream and downstream distribution functions.
The simulated (dashed lines) and analytical (solid lines) distribution
functions are shown in Fig.~\ref{densities}b, for $\xh=15$ (upstream,
red) and $\xh=-15$ (downstream, blue), respectively. Since $\phih>0$
at these locations, the flow velocity of the bulk ion distribution is
above the far upstream value, $-\Vh$. Analytically, the velocity space
is empty above $\vh_{-}=-2.29$ in the downstream region, there is a
reflected population with a cutoff at $\vh_{+}=2.53$ in the upstream
region. Similar cutoffs are also seen in the simulation; in
particular, there is practically no ion population in the
trapped region downstream, which supports our choice of ion
ansatz.

\begin{figure}
\includegraphics[width=0.45\textwidth]{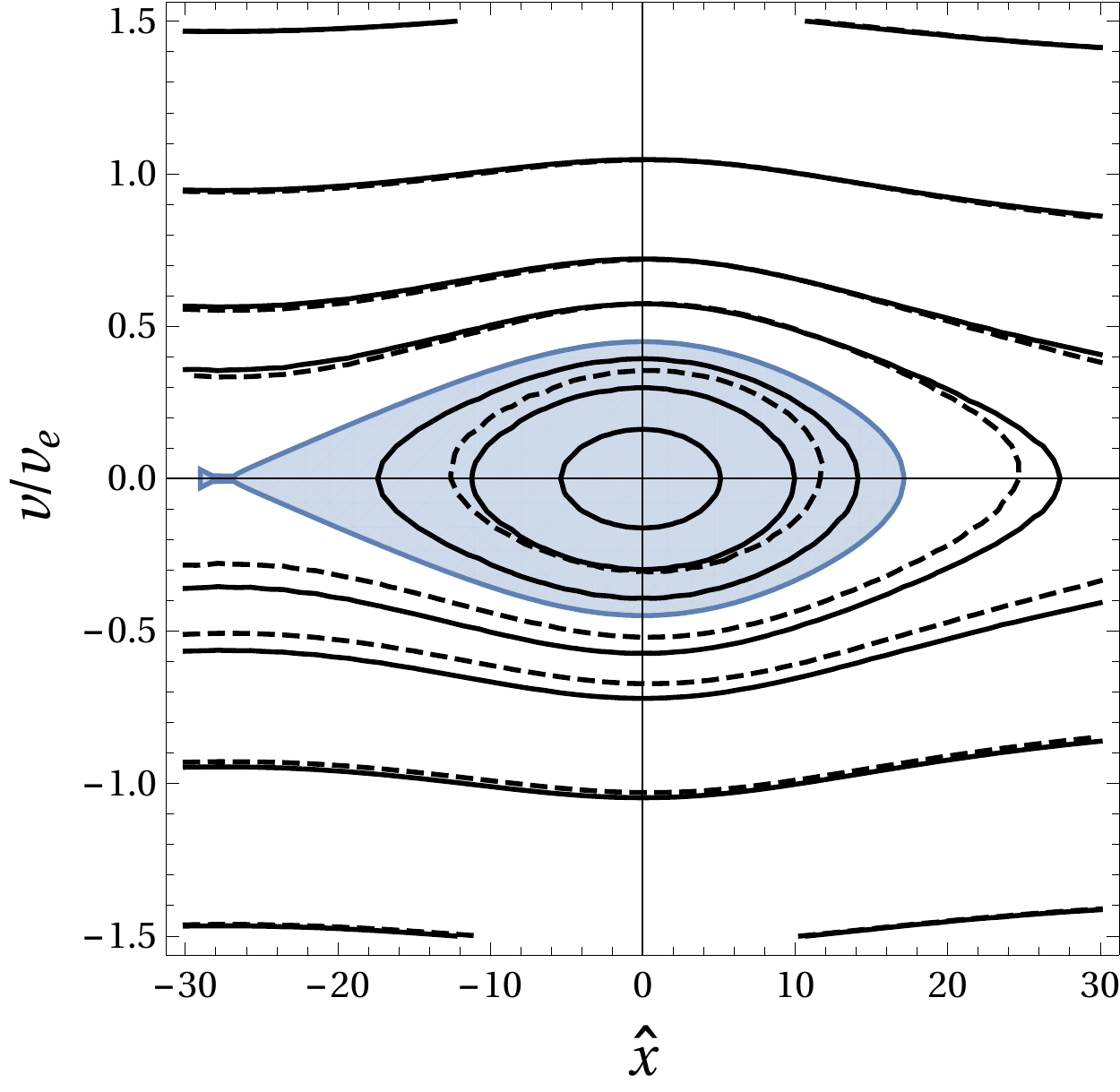}
\caption{\label{electrondist1} Contours of the electron distribution
  function, normalized such that
  $\tilde{f}_e(\xh,v/v_e=0)=\hat{n}_e(\xh)$.  Solid lines: assuming
  Maxwell-Boltzmann electron distribution; dashed lines: simulation
  result. The lines show
  $\tilde{f}_e(\xh,v/v_e)=\{0.4,\,0.75,\,1,\,1.2,\,1.24,\,1.28\}$,
  converging about the origin in this order (note that for the
  simulation results, the $1.24$ and $1.28$ contours do not
  exist). Blue shaded area shows the trapped region in the electron
  phase space.  }
\end{figure}

\begin{figure}
\includegraphics[width=0.45\textwidth]{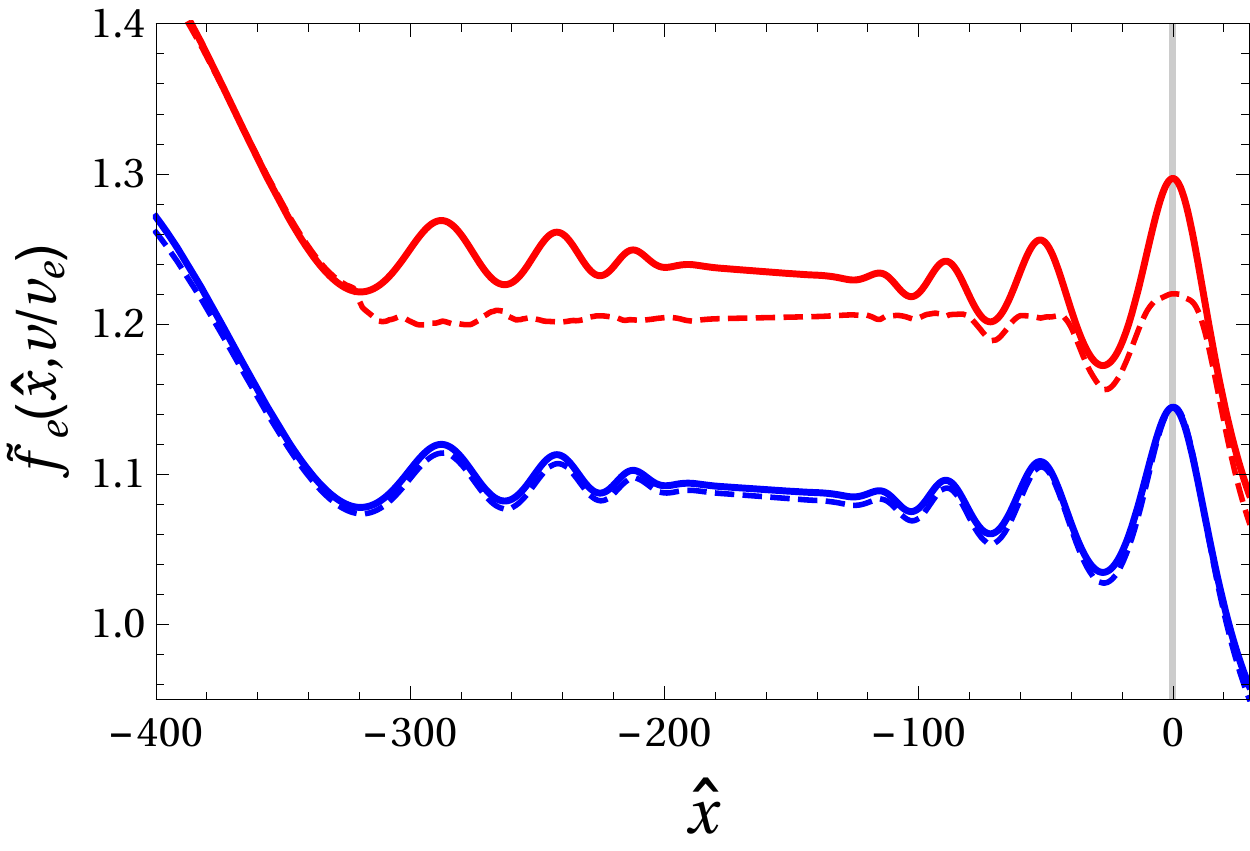}
\caption{\label{electrondist2} Electron distribution function,
  normalized such that
  $\tilde{f}_e(\xh,v/v_e=0)=\hat{n}_e(\xh)$. Solid lines: assuming
  Maxwell-Boltzmann electron response; dashed lines: simulation
  result. Red: $\tilde{f}_e(\xh,v/v_e=0)$; blue:
  $\tilde{f}_e(\xh,v/v_e=+0.5)$.  }
\end{figure}

Finally, we consider the electron distribution, which in the
analytical model is taken to be Maxwell-Boltzmann. However, in
reality, electrons can get trapped around potential maxima, and there
they will develop deviations from a Boltzmann response. This situation
indeed happens in the simulation, as illustrated in
Figs.~\ref{electrondist1} and \ref{electrondist2}. The trapped region
in the phase space is illustrated by the blue shaded area in
Fig.~\ref{electrondist1}, where the kinetic energy of electrons is
lower than the potential difference between $\phimh$, and $\phih_{\rm
  min,-1}$, the first potential minimum downstream. When we compare
the Maxwell-Boltzmann model (solid lines), and the simulation results
(dashed), we find that they are close at high electron speeds, where
the potential can only slightly perturb the total energy of the
electrons. However, at lower speeds the simulations show increasing
deviations from the Maxwell-Boltzmann response. In particular, the
distribution inside the trapped region is much flatter (thus the two
highest contours do not exist for the simulated distribution). When we
take a velocity $|v/v_e|>0.45$, where no trapped region exists, we see
a remarkable agreement between the analytical and simulated electron
distribution functions, as shown by the blue lines in
Fig.~\ref{electrondist2} ($v/v_e=+0.5$), while in all the trapped
regions about the local $\phih$ maxima, we find a flattening of the
distribution, and a reduced electron density, accordingly, see the red
lines taken at $v/v_e=0$.

\section{Shock solutions and reflected ions}
\label{sec:secthree}
The Vlasov-Poisson simulations presented in the previous section
confirm that the semi-analytical model captures the main properties of
the shock structure. We therefore proceed to use the model to study
the shock solutions, in particular the effect of impurities and
electron-to-ion temperature ratio.

To determine the maximum electrostatic potential, we solve $
\Phi(\phimh, \phimh)= 0$, with input parameters $\tau$ and $\Vh$. As
it was pointed out in Ref.~\citenum{cairns}, not all combinations of
$\tau$ and $\Vh$ give solutions. In particular, numerical results
indicate that $\tau$ needs to exceed a specific value for the
existence of solutions. As in Ref.~\citenum{cairns}, for a single ion
species (protons), we have a solution for $\Vh=4.5$ and $\tau=15$
(corresponding to $M=1.16$) with a maximum normalized electrostatic
potential $\phimh=1.29$ and a correspondingly reflected fraction
$\alpha_i=0.0019$.

We start by adding a small amount of impurities to the pure proton
plasma. Already, the addition of only $1\%$ of a fully ionized carbon
species ($n_z/n_i=0.01$) changes the solutions considerably.  The
maximum normalized electrostatic potential increases to
$\phimh=1.676$, and consequently, the reflected main ion fraction
doubles to $\alpha_i=0.0038$.  As the fraction of carbon impurities
increases, $\phimh$ continues to increase; however, for more than
$9\%$ carbon, there exists no solutions for the shock velocity
$\Vh=4.5$ and temperature ratio $\tau=15$.

\begin{figure*}
\includegraphics[width=0.45\textwidth]{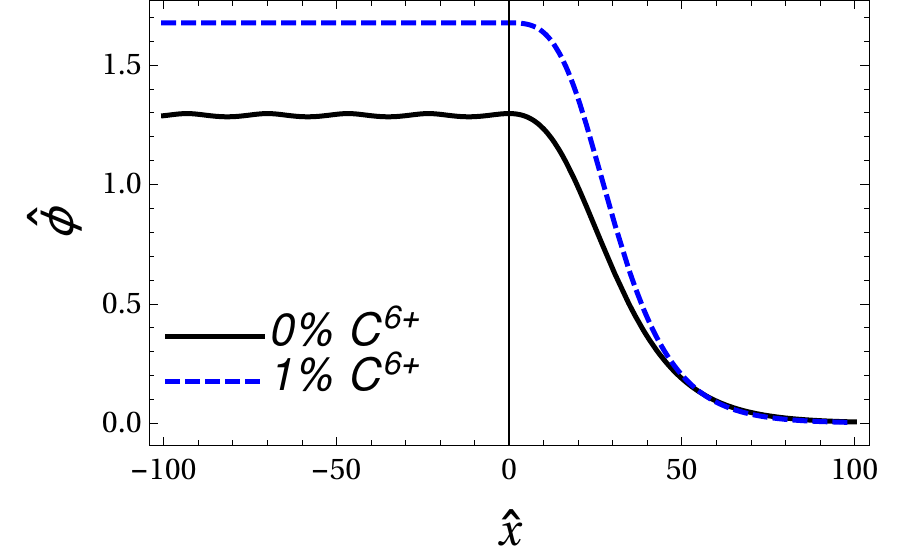}
\put(-30,95){\Large (a)}  
\put(160,95){\Large (b)}
$\quad$
\includegraphics[width=0.44\textwidth]{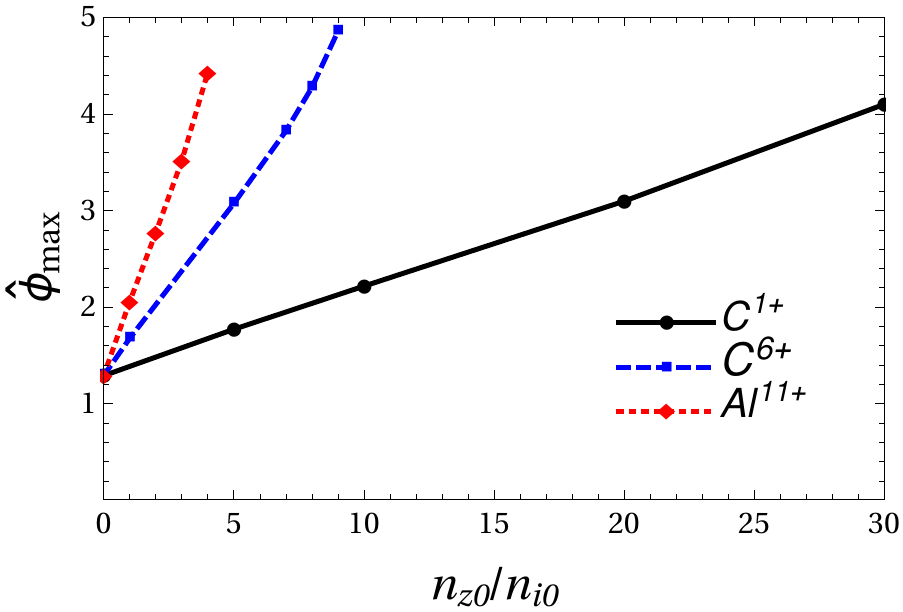}
\caption{\label{potentials}(a) Electrostatic potential for $\Vh=4.5$ and
  $\tau=15$ for the case of a pure proton plasma (solid) and a $1\%$  of fully
  ionized carbon (dashed). (b) Maximum normalized electrostatic
  potential for various fractions of impurities for $\Vh=4.5$ and
  $\tau=15$.}
\end{figure*}

\begin{figure*}
\includegraphics[width=0.45\textwidth]{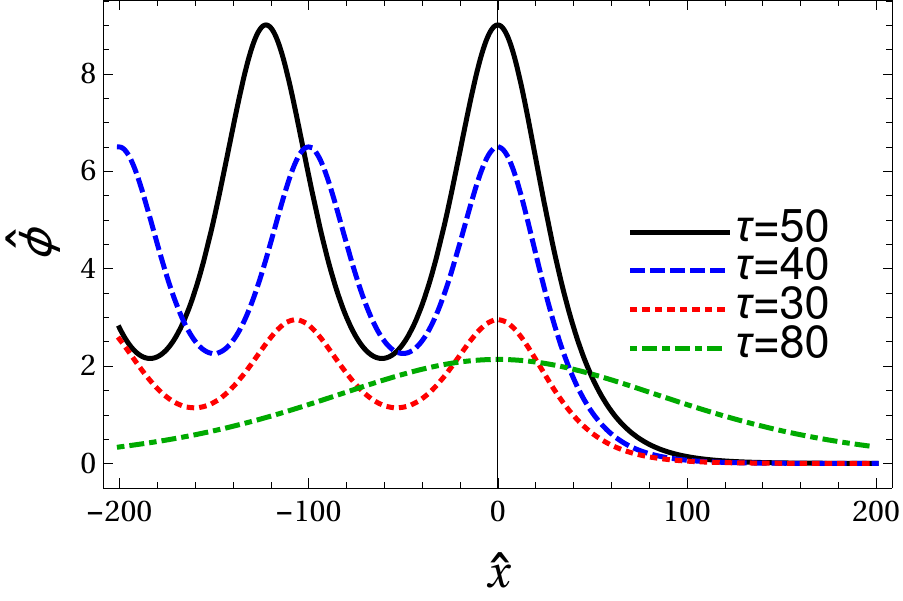}
\put(-35,95){\Large (a)}  
\put(160,95){\Large (b)}
$\quad$
\includegraphics[width=0.45\textwidth]{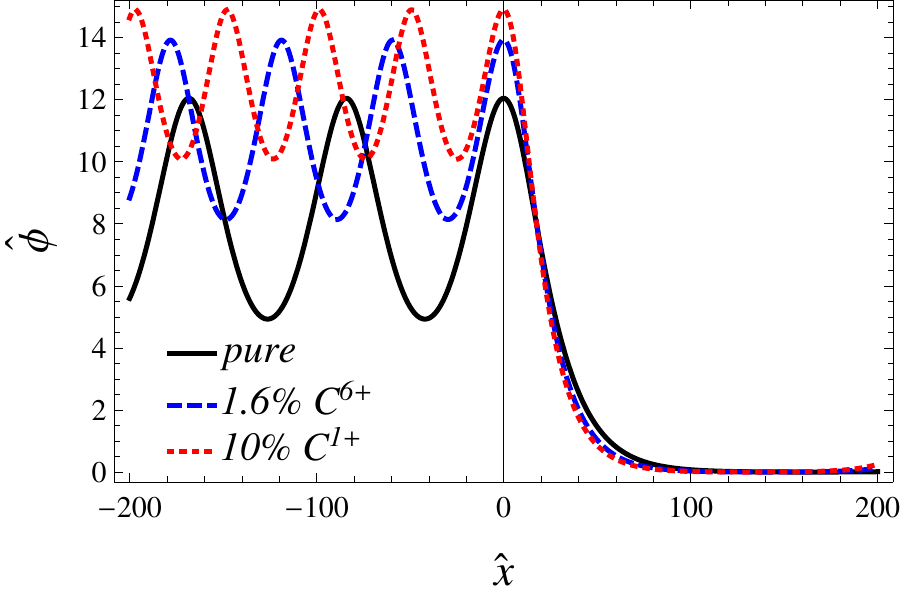}
\caption{\label{hightaupotentials}(a) Electrostatic potential for
  various temperature ratios $\tau$, for the case of a pure proton plasma
  and Mach number close to unity. Solid is $\Vh=7.8$ and $\tau=50$, blue
  dashed is for $\Vh=7$ and $\tau=40$, red dotted is for $\Vh=6$ and
  $\tau=30$, green dash-dotted is for $\Vh=9.2$ and $\tau=80$. (b)
  Electrostatic potential for $\Vh=8$ and $\tau=50$, for the case of a pure
  proton plasma (solid), $1.6\%$ of fully ionized carbon (blue
  dashed) and $10\%$ of singly ionized carbon (red dotted).}
\end{figure*}

The existence of a shock solution is also affected by the ionization
degree. We find that for singly ionized carbon, solutions exist even
for very high concentrations, up to $30\%$.  Figure \ref{potentials}a
shows the electrostatic potential for $\Vh=4.5$ and $\tau=15$ in the
case of a pure hydrogen plasma (solid) and a hydrogen plasma
contaminated by $1\%$ of fully ionized carbon (dashed).  Figure
\ref{potentials}b shows the dependence of the maximum normalized
electrostatic potential for three types of impurities on the impurity
fraction. We find that for a given impurity, $\phimh$ increases
  approximately linearly with impurity concentration. In addition,
  $\phimh$ increases more rapidly with concentration for impurities
  that have a higher charge, however the slope is not simply
  proportional to the charge number of the impurity (which is evident
  from comparing $\phimh$ for $1\%$ $\rm C^{6+}$ with $6\%$ $\rm
  C^{1+}$). 

It is instructive to investigate how the solutions change if the
temperature ratio $\tau$ is assumed to be larger. Higher $\tau$ is
relevant for high-intensity laser-driven ion acceleration experiments
as the laser mostly heats the electrons.  Note that the Mach number is
reduced with growing $\tau$, but the reflected ion fraction, which
involves only $\Vh$ and $\phimh$, will not be directly affected by
the change in $\tau$, only indirectly.  In agreement with
Ref.~\citenum{CairnsPPCF}, we find that for higher $\tau$ the range of
possible Mach numbers is wider.

Figure~\ref{hightaupotentials}(a) shows the electrostatic potential
for various values of $\tau$. The maximum value increases with $\tau$
until a certain limit is reached (at around $\tau=60$) and above this
value, the solutions become symmetric and soliton-like. The presence
of impurities increases the maximum electrostatic potential and
thereby the reflected ion fraction. A similar effect results from the
presence of $1.6\%$ of fully ionized carbon as $10\%$ of singly ionized
carbon, as shown in Figure~\ref{hightaupotentials}(b).

\begin{figure*}
\includegraphics[width=0.45\textwidth]{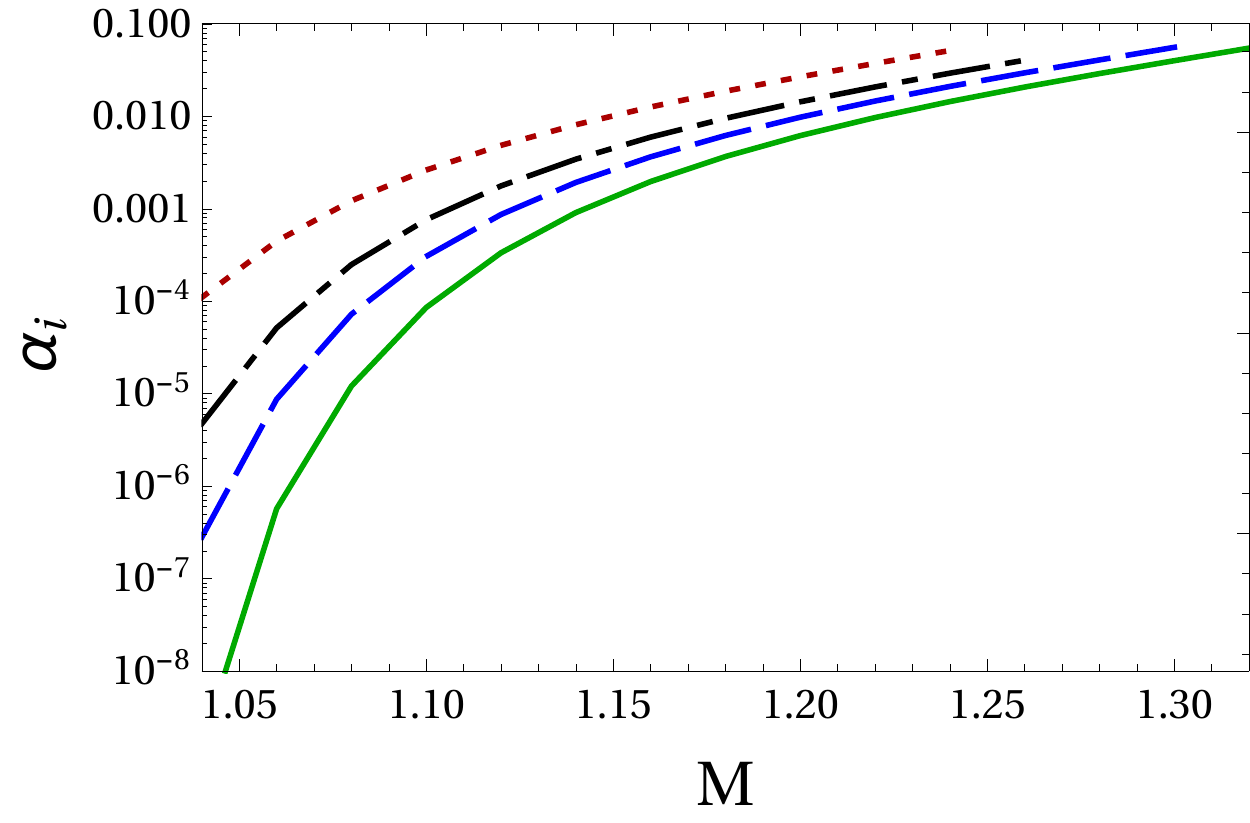} 
\put(-30,40){\Large (a)}  
\put(160,40){\Large (b)}
$\quad$
\includegraphics[width=0.45\textwidth]{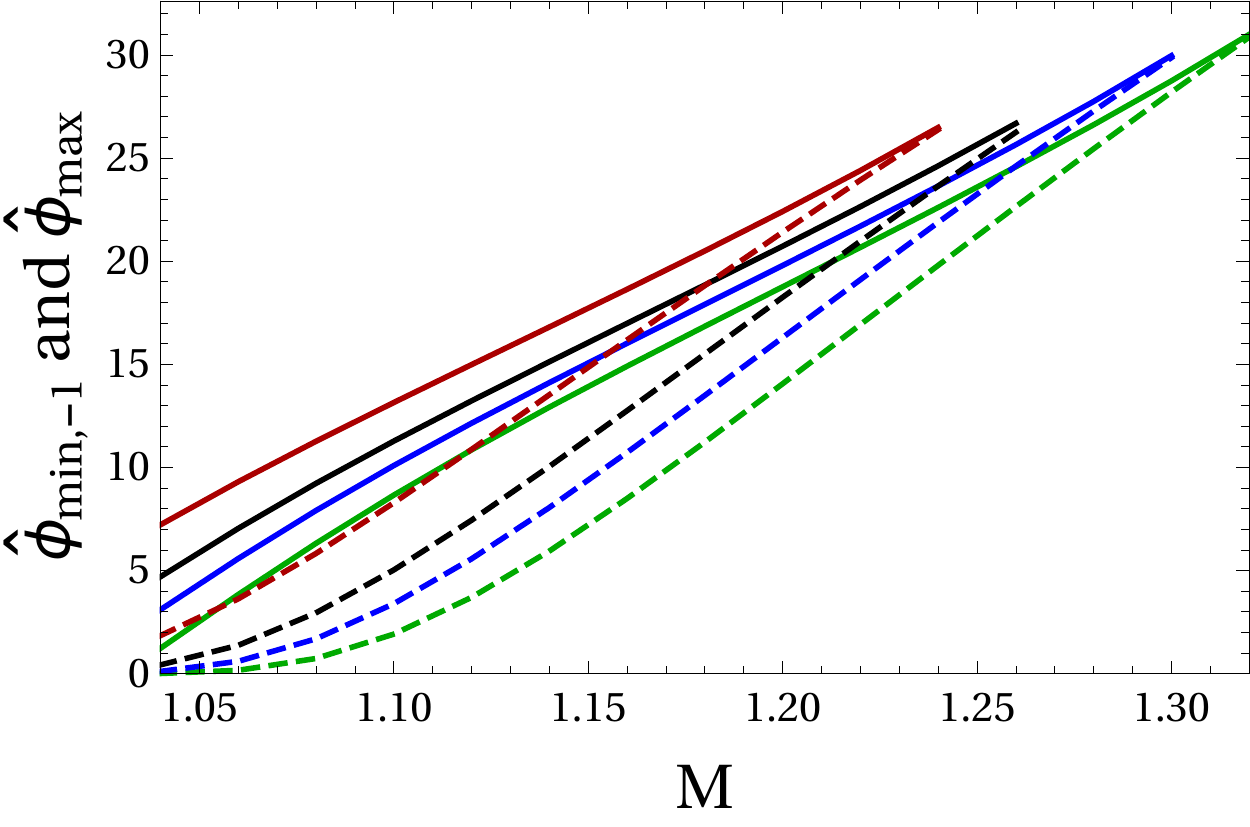}
\caption{\label{alfafig} (a) Reflected ion fraction as function of
  Mach number for various concentrations of fully ionized carbon in a
  hydrogen plasma. Solid (green) line is for pure plasma, dashed
  (blue) is for $1\%$ carbon, dash-dotted (black) is for $2\%$ carbon and
  dotted (dark red) is for $4\%$ carbon. The electron-to-ion temperature
  ratio is $\tau=50$. (b) Maximum (solid) and minimum (dashed) of the
  electrostatic potential in the downstream region for the impurity
  concentrations given in (a). The colors represent different
  concentrations of fully ionized carbon in a hydrogen plasma. Green,
  blue, black, and dark-red correspond to $0\%$, $1\%$, $2\%$, and $4\%$,
  respectively. }
\end{figure*}

We find that the reflected ion fraction grows with Mach number for
fixed electron-to-ion temperature ratio. Figure~\ref{alfafig}a shows
the growth of the reflected ion fraction for various fractions of
fully ionized carbon in a hydrogen plasma. Note that a small change in
the Mach number can lead to several orders of magnitude of change in
the reflected ion fraction. Figure~\ref{alfafig}b shows that the
difference between the maximum and minimum of the electrostatic
potential in the downstream region is smaller with increasing impurity
concentration, and at a certain shock speed, solutions cease to
exist. This maximum shock speed decreases with impurity
concentration.

It is interesting to point out that hydrogen has a charge to mass
ratio $Z_i/\mih=1$, that is approximately twice as large as that of
any other fully charged ion species. This difference leads
to a strong disparity between the behavior of hydrogen as main ion or
as impurity, which we would like to illuminate mathematically in the
following. Besides $\tau \gg 1 $, we only assume that the impurity
concentration is sufficiently small for the propagation speed of the
shock to be determined by the main ions. This assumption means that
$\Vh=M \sqrt{Z_i\tau/\mih}$ with some $M\gtrsim 1$. We can express
Eq.~(\ref{alpha}) in terms of $M$ by using the previous expression
together with $\phimh=F M^2\tau/2$, where the constant $F$ can be
determined by solving $\Phi(\phimh,\phimh)=0$. Furthermore, for large
$\tau$, the $\erf (\tilde{V}_j )$ terms in Eq.~(\ref{alpha})
for $j=i$ can be very well approximated by $1$, and recalling that
$\Tih=1$, we find that the reflected fraction of main ions is
\begin{equation}
  \alpha_i= \frac{1}{2}\left\{ 1- \erf
  \left[(1-\sqrt{F})M\sqrt{\frac{Z_i \tau}{2}}\right] \right\}.
\label{alphai}
\end{equation}  
Now it is apparent that $F=1$ gives $\alpha_i=1/2$, and $\alpha_i$
decreases rapidly as $F$ gets smaller than 1. In particular, if
$1-\sqrt{F}$ becomes larger than $\sqrt{2/(Z_i\tau)} \ll 1$,
$\alpha_i$ is very small. In practice, this result sets $\phimh$ to be
slightly smaller than $ M^2\tau/2$ for physically interesting cases
(i.e., where $\alpha_j$ is not vanishingly small). Using the
above procedure, we find the following expression for the reflected
fraction of impurities
\begin{equation}
  \alpha_z= \frac{1}{2}\left\{ 1+\erf \left[M\sqrt{\frac{F Z_z
        \tau}{2\Tzh}}\left(1-\sqrt{\frac{\mzh Z_i }{F\mih
        Z_z}}\right)\right] \right\}.
\label{alphaz}
\end{equation} 
Since $F\approx 1$, and for hydrogen main species $\mzh Z_i/(\mih
Z_z)>1$, the term in the parentheses is a negative number, and
$\alpha_z$ evaluates to an extremely small number. However, for
hydrogen impurity the parenthetical term is positive, and $\alpha_z$
is close to unity, although $1-\alpha_z$ is not necessarily
extremely small. In other words, in a hydrogen plasma, the impurities
are practically unreflected by the shock. However, in a non-hydrogenic
plasma almost all hydrogen impurities are reflected. The latter
conclusion is illustrated in the following through a numerical
simulation.

We performed a \gkeyll~simulation for a fully ionized aluminum plasma
($Z_{i}=13$, $\mih=27$) with hydrogen impurity ($Z_z=1$, $\mzh=1$) of
concentration $n_z/n_i=0.01$, and equal ion temperatures
$T_i=T_z$. The normalized electron temperature is $\tau=45$, and the
simulation is initiated with a density discontinuity with a density
ratio $2$.  The simulation used similar resolution parameters to those
in the simulation shown in Sec.~\ref{sec:sectwo}, except that the
range of velocities is $\left\{-18 v_i,\,54 v_i\right\}$ for aluminum
and $\left\{-6 v_z,\,18 v_z\right\}$ for hydrogen, and the number of
cells in velocity space is $96$.

We consider the solution at $t \sqrt{m_e/m_p}\omega_{pe}=35$, where
$\omega_{pe}=\sqrt{e^2 n_{e0}/(\epsilon_0 m_e)}$, with $n_{e0}$ the
far upstream electron density. In the simulation we find that the
shock potential is $\phimh=24.09$, and the shock propagates with a
speed $\Vh=5.72$, corresponding to $M=1.229$. According to $\phimh=F
M^2\tau/2$, this set of parameters translates to $F=0.709$, which is rather low,
suggesting that the reflected main ion fraction is small. Indeed, from
Eq.~(\ref{alphai}) we estimate $\alpha_i=1.33 \cdot 10^{-6}$, and in
the simulation $\alpha_i$ is so small that it cannot be meaningfully
evaluated within the finite numerical accuracy.  However, as expected,
the ratio of reflected hydrogen impurities is substantial: from
Eq.~(\ref{alphaz}) we estimate $\alpha_z=0.889$, and taking the ratio
of the densities in the incoming and reflected populations at
$\xh=62.6$ yields a comparable $\alpha_z\approx 0.874$. Finally, we
note that for the same $\tau$ and $\Vh$ the semi-analytical
calculation gives a somewhat higher shock potential $\phimh=25.953$
($F=0.763$) that corresponds to $\alpha_i=8.69 \cdot 10^{-5}$ (note
the sensitivity to $F$) and $\alpha_z\approx 0.93$.

\begin{figure*}
\includegraphics[width=0.4\textwidth]{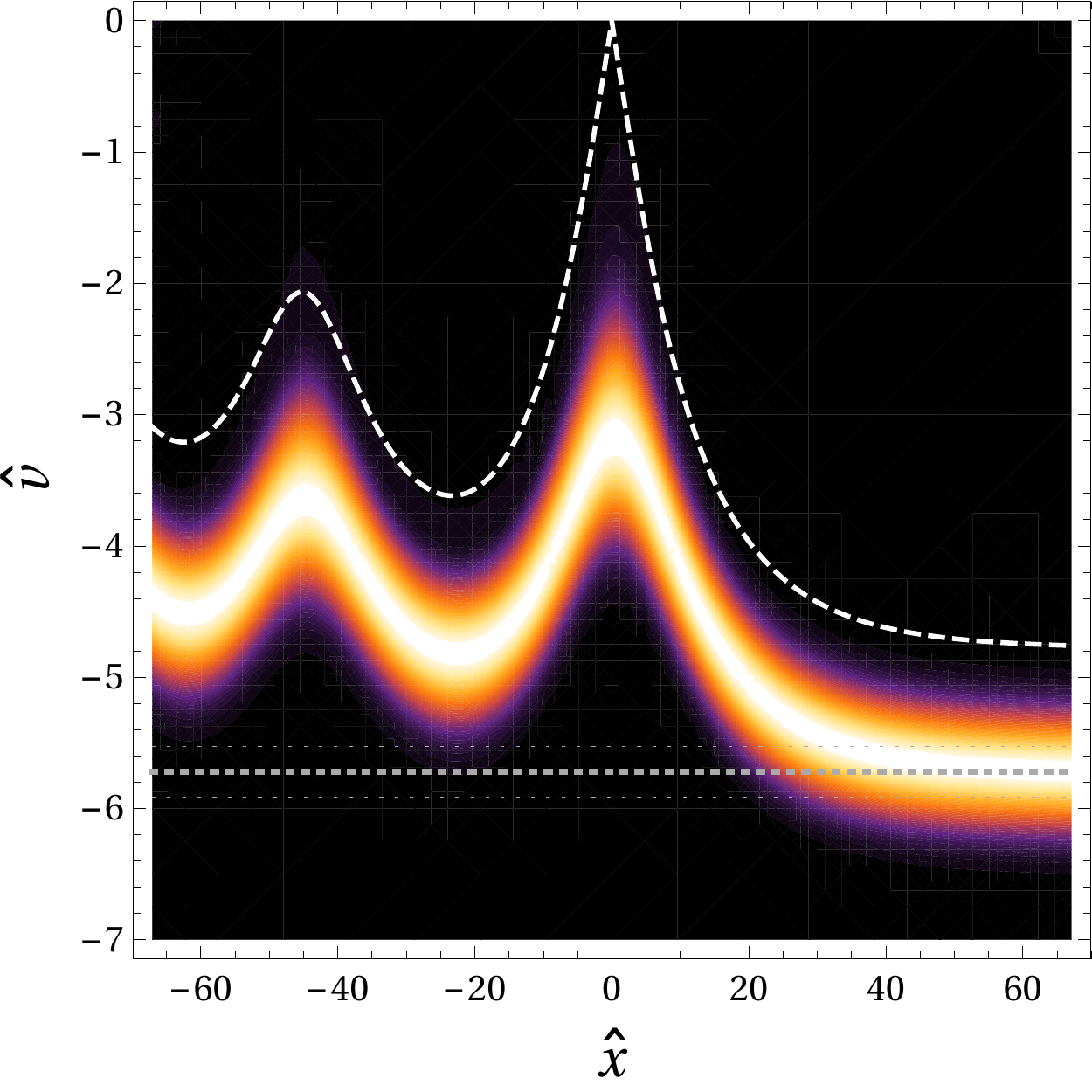} 
\put(-130,140){\large \color{white} (a)}  
$\quad$
\includegraphics[width=0.4\textwidth]{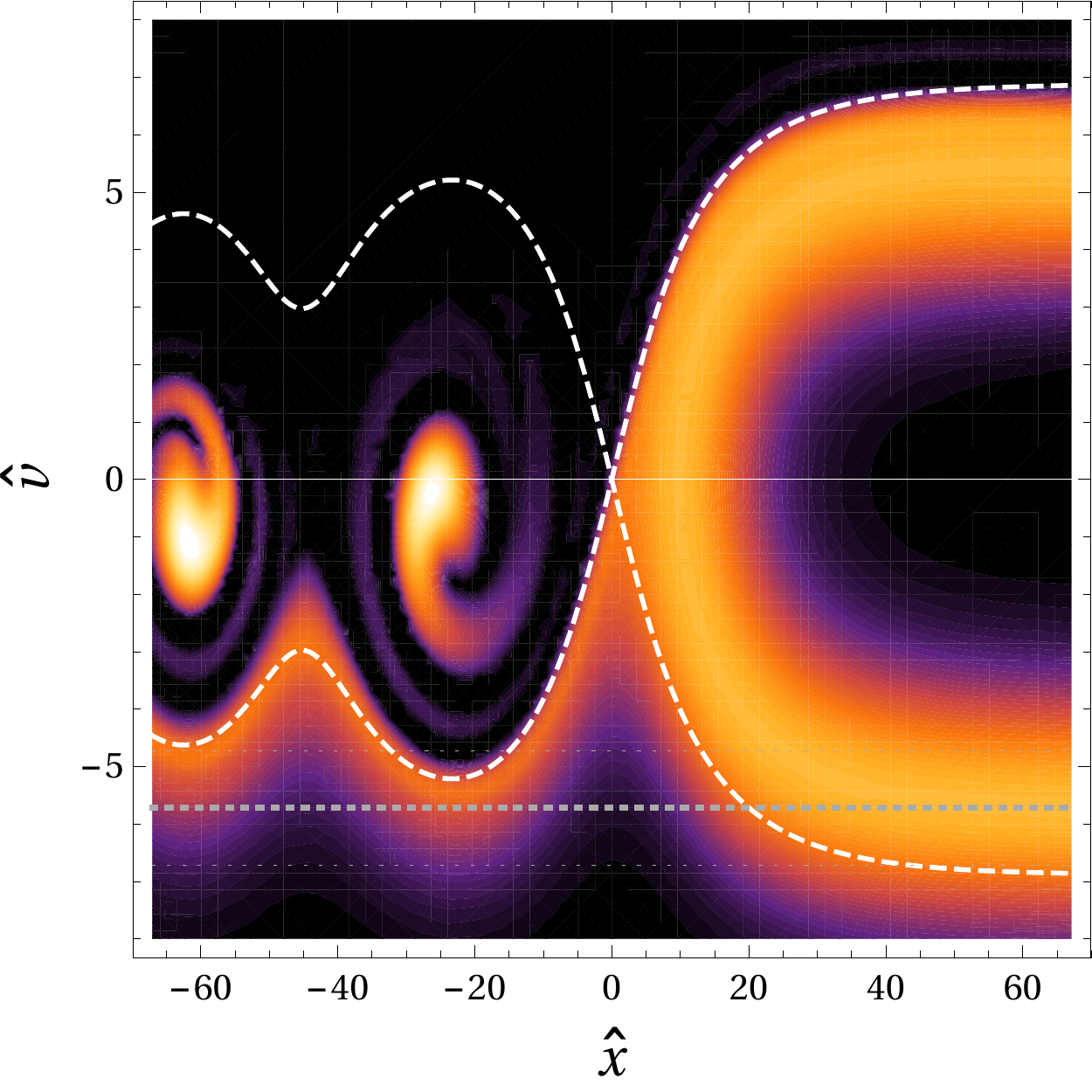}
\put(-130,140){\large \color{white} (b)}
$\qquad\quad$
\includegraphics[width=0.07\textwidth]{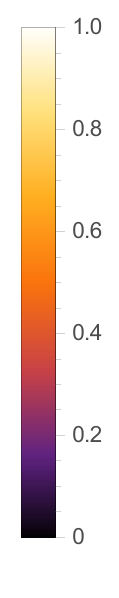}
\put(-60,145){\large $\sqrt{\hat{f}/\text{max}(\hat{f})}$}
\caption{\label{distfigs} Distribution functions of the aluminum main
  ion species (a), and the hydrogen impurity (b), from a \gkeyll~
  simulation. Dashed lines correspond to the separatrices,
  $\vh_{\pm}(\xh)$. The far-upstream ion flow speed in the shock
  frame, $-\Vh$, is shown with thick dotted line, and one thermal
  speed width of the distribution is indicated by the thin dotted
  lines $-\Vh\pm v_j$. Mind the different $\vh$ scales.  }
\end{figure*}

\begin{figure} 
\includegraphics[width=0.4\textwidth]{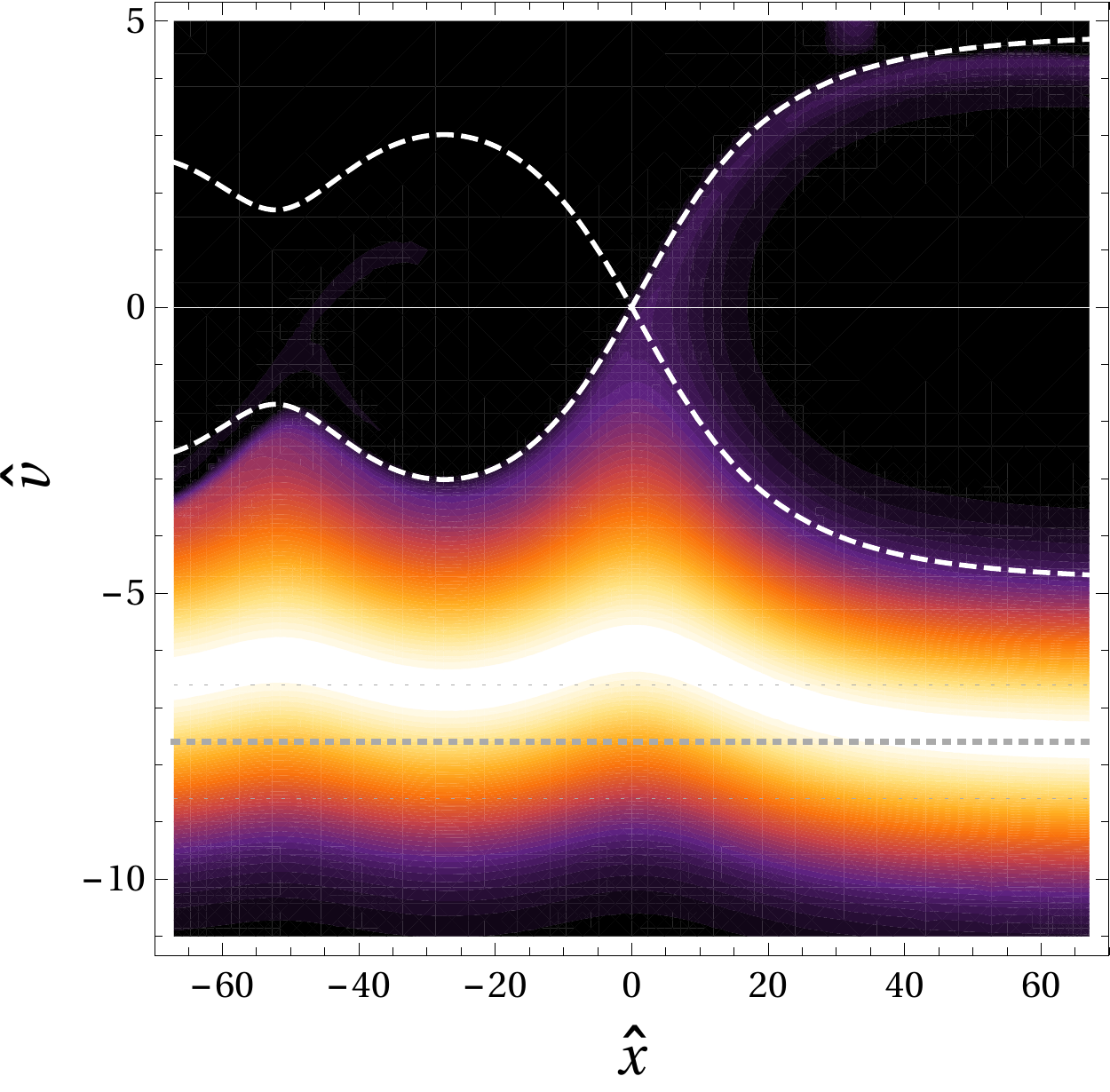} 
\caption{\label{purehdist} Ion distribution function from a pure
  hydrogen plasma simulation. Dashed lines correspond to the
  separatrices, $\vh_{\pm}(\xh)$. The far-upstream ion flow
  speed in the shock frame, $-\Vh$, is shown with thick dotted line,
  and one thermal speed width of the distribution is indicated by the
  thin dotted lines $-\Vh\pm v_j$. }
\end{figure}

Sections of the simulated distribution functions of the two ion
species around the shock front are shown in Fig.~\ref{distfigs}. In
addition, the separatrices $\vh=\vh_{\pm}(\xh)\equiv\pm
\sqrt{2(Z_j/\mjh)[\phimh-\phih(\xh)]}$, with $\phih(\xh)$ also taken
from the simulation, are shown with dashed lines.  The far upstream
flow speed $-\Vh$, together with species thermal speed deviations from
it, $-\Vh\pm v_j$, are indicated by the thick and thin dotted lines,
respectively. For aluminum, shown in Fig.~\ref{distfigs}a, we can see
that the maximum of the far upstream ion distribution is several
thermal speeds below the separatrix, which is consistent with the low
reflected ion fraction. For hydrogen, shown in Fig.~\ref{distfigs}b, a
major part of the distribution is above the separatrix, hence the
large reflected fraction. In contrast to the pure hydrogen plasma case
shown earlier, and the aluminum main species, the phase space of the
hydrogen impurity is not empty behind the shock, above the separatrix:
blobs of hydrogen are being trapped in these regions. Interestingly
$f_z$ reaches twice as high values in the trapped region than
upstream, suggesting that they cannot originate from the upstream
region. In fact, the blobs are torn away from the large density region
of the initial density discontinuity, and are now trapped by the
downstream oscillations of the potential.

For comparison, in Fig.~\ref{purehdist} we also show the ion
distribution in the pure hydrogen simulation of Sec.~\ref{sec:sectwo},
where most of the far-upstream distribution is below the separatrix,
thus only a small fraction of ions is reflected.

\section{Discussion and conclusions}
\label{sec:secfour}
 In this paper, we extend the analysis of low-Mach number
 electrostatic shock structures with a semi-analytical model. We
 assume Maxwell-Boltzmann electrons and an ion distribution that is
 extended along contours of constant total energy from a Maxwellian
 far upstream. The self-consistent electrostatic field is calculated
 using Poisson's equation. Regarding these aspects, it is similar to
 the model of Cairns et al~\cite{cairns,CairnsPPCF}. However, in the
 model described in Refs.~\citenum{cairns} and \citenum{CairnsPPCF}, the ion
 distribution in the downstream region is finite below $v=0$ in the
 shock frame, and zero above, which appears to be inconsistent in a
 steady state shock model (except in monotonic shocks without
 trapped regions downstream).  The difference between the models
 affects only the downstream properties of the shock but not its
 existence or the reflected ion fractions.

We compare the semi-analytical model to Eulerian Vlasov-Poisson
simulations with \gkeyll, where we consider shocks generated by the
decay of an initial density discontinuity. We find that the model well
reproduces the simulated shock potential and Mach number. The electron
distribution function is well approximated by a Maxwell-Boltzmann
distribution in most of the phase space, while simulations show
signatures of trapping in the downstream oscillations of the
potential, depleting the low speed population of electrons around the
local potential maxima. In single species and hydrogen bulk
simulations the ion distribution is well captured by our ansatz; in
particular, the downstream trapped regions of the ion phase
space are empty. However, for non-hydrogenic main species, trapping of
hydrogen impurities can occur.

We used the semi-analytical model to study the effect of heavy
ion impurities on the shock parameters and reflected ion
fraction. This study is of relevance to laser-based ion acceleration
experiments, which are rarely free of impurities, for
  example. We find, that only a few percent of fully ionized carbon
impurity in a hydrogen plasma will give a large effect on the
reflected ion fraction. This is because the maximum electrostatic
potential grows considerably. However, the maximum reflected fraction
of hydrogen ions remains below 10\% and the Mach number is low,
therefore these shocks are not expected to be efficient accelerators
of a large number of ions to very high energies, rather their strength
lies in their mono-energeticity.

We provide accurate analytical expressions for the reflected fractions
of main ions and impurities, which illuminate the different behavior
of hydrogen, depending on its role as main ion or impurity. In a
hydrogen plasma with a large electron-to-ion temperature ratio, the
reflection of non-hydrogenic impurities is vanishingly small. On the
other hand, one way to increase the reflected fraction of hydrogen
ions is to create a shock in a heavy ion plasma containing hydrogen
impurities, in which case almost all of the hydrogen will be
reflected, with a speed close to twice the shock speed. Similarly, the
different behavior of hydrogen as a main species or as an impurity has
been reported before in the context of the expansion of a
  multi-species plasma into vacuum \cite{tikhonchuk,robinson}.

\acknowledgments The authors would like to thank A~Stahl for numerical
advice, together with L~Gremillet, E~Siminos, T~C~DuBois,
  and A~Sundstr\"{o}m for useful comments on the manuscript. This
work was supported by the International Career Grant
(Dnr.~330-2014-6313) from Vetenskapsr{\aa}det, and Marie Sklodowska
Curie Actions, Cofund, Project INCA 600398; the European Research
Council (ERC-2014-CoG grant 647121), the Knut and Alice Wallenberg
Foundation, and the National Science Foundation (NSF) SHINE award
No.~AGS-1622306. This development of \gkeyll~code is partly funded by
the U.S.~Department of Energy under Contract No.~DE-AC02-09CH11466 and
by the Air Force Office of Scientific Research under grant number
FA9550-15-1-0193. This work used the Extreme Science and Engineering
Discovery Environment (XSEDE), which is supported by NSF grant number
ACI-1548562.

\bibliography{ShockEquations_rev1.bib} 

\end{document}